\documentclass[twocolumn]{aastex631}

\usepackage{color}
\usepackage{amsmath}
\usepackage{changepage}
\usepackage{float}

\begin{document}
\nolinenumbers


\title{Modeling Realistic Heating Profiles of Transition Region Hot Loops on the Sun: Evidence for Impulsive Heating and Non-equilibrium Ionization}

\correspondingauthor{Shah Mohammad Bahauddin}
\email{shahmohammad.bahauddin@colorado.edu}

\author{Shah Mohammad Bahauddin}
\affiliation{Laboratory for Atmospheric and Space Physics, University of Colorado, Boulder, CO 80303, USA}

\author{Stephen J Bradshaw}
\affiliation{Department of Physics and Astronomy, Rice University, Houston, TX 77005, USA}

\begin{abstract}

The study examines the heating profile of hot solar transition region loops, particularly focusing on transient brightenings observed in IRIS 1400\AA{} slit-jaw images. The findings challenge the adequacy of simplistic, singular heating mechanisms, revealing that the heating is temporally impulsive and requires a spatially complex profile with multiple heating scales. A forward modeling code is utilized to generate synthetic IRIS emission spectra of these loops based on HYDRAD output, confirming that emitting ions are out of equilibrium. The modeling further indicates that density-dependent dielectronic recombination rates must be included to reproduce the observed line ratios. Collectively, this evidence substantiates that the loops are subject to impulsive heating and that the components of the transiently brightened plasma are driven far from thermal equilibrium. Heating events such as these are ubiquitous in the transition region and the analysis described above provides a robust observational diagnostic tool for characterizing the plasma.

\end{abstract}


\section{Introduction} \label{sec:intro}

The dramatically increased spatial and temporal resolution of the current generation of observing instruments has revealed that, besides large-scale impulsive events such as solar flares, small-scale dynamic events that evolve in a time-scale of several seconds to several minutes can release bursts of substantial heating ~\citep[e.g.,][]{2011ApJ...730L...4J, 2012SSRv..169..181T, 2012ApJ...750L..25J, 2013Natur.493..501C, 2014ApJ...787L..10W, 2016GMS...216..431V, 2019ApJ...887...56T, 2021NatAs...5..237B}. Although widespread, these impulsive heating events have eluded detection in earlier-generation instruments, all the while potentially contributing significantly to the heating of the solar corona ~\citep[e.g.,][]{2014ApJ...787L..10W, 2014Sci...346E.315H, 2015RSPTA.37340269D, 2016ApJ...826L..18B, 2016AGUFMSH31B2577W, 2021NatAs...5..237B, 2022ApJ...926...52D, 2022FrASS...920116A, 2023ApJ...945..143C}. One of the potential mechanisms at this scale involves the transformation of magnetic energy into thermal energy and bulk kinetic energy through viscous heating at small spatial scales. This heating mechanism, estimated to release energy on the order of $10^{24}$ erg~\citep{1957PhRv..107..830P, 1958IAUS....6..123S, 1983NASCP.2280..23P}, is assumed to be one of the basic units of impulsive energy release and termed nanoflares ~\citep{1983NASCP.2280..23P, 1988ApJ...330..474P, 1988sscd.conf....2P}. Subsequent theoretical analysis ~\citep[e.g.,][]{1994ApJ...422..381C, 1997ApJ...478..799C, 2004ApJ...605..911C} and recent observations suggest that the solar corona may be heated to such high temperatures by swarms of nanoflares ~\citep[e.g.,][]{2005ApJ...618.1020G, 2008ApJ...682.1351K, 2014ApJ...784...49C, 2015ApJ...811..106H, 2016ApJ...833..217B, 2017ApJ...834...10R} with each event having a spatial scale of approximately $10^3$ km. In reality, it is probable that these events vary in spatial scale and characteristic energies.

Given the small spatial scale and transient nature of nanoflares, directly imaging their dynamics has remained exceedingly challenging, necessitating the search for indirect evidence of their existence ~\citep[e.g.,][]{2001ApJ...553..440K, 2006A&A...458..987B, 2011ApJS..194...26B, 2015RSPTA.37340260C}. Despite the difficulties of directly observing nanoflares, there is ample observational evidence for transient brightenings in coronal structures, such as in coronal loops ~\citep[e.g.,][and references therein]{2002ApJ...570L.105W, 2014ApJ...787L..10W, 2013ApJ...771...21W, 2013ApJ...770L...1T, 2014Sci...346E.315H, 2016ApJ...826L..18B, 2016ApJ...827...99T, 2017MNRAS.464.1753H, 2021SoPh..296...84D}, that can be associated with nanoflare-like impulsive heating. Recent enhancements in the spatial and temporal resolution of observations due to the Interface Region Imaging Spectrograph (IRIS, ~\citep{2014SoPh..289.2733D}) have made it possible to study these brightenings in rich detail ~\citep[e.g.,][and references therein]{2021SoPh..296...84D} and opened the possibility to unravel their origin. One of such multi-channel multi-instrument observational analysis made by the authors of this paper can be found in ~\cite{2021NatAs...5..237B} where we have shown that small low-lying loops residing in the transition region are multi-stranded and release bursts of energy via reconnection mediated heating. This paper aims to investigate the heating profile and the resulting hydrodynamic evolution of these loops and critically investigate the populations of the emitting ions to recreate the observed spectra.

Models of small, low-lying coronal loops typically adopt a dynamic ``hot" loop framework where the temperature profile undergoes a rapid increase within the transition region before leveling off in the corona. Thermal conduction plays a vital role in the energy dynamics of these loops, with a significant portion of the heat dissipated within the corona, while the remainder is transported to the upper chromosphere and transition region where radiation is more efficient. Hydrodynamically, these loops maintain nearly constant pressure and consequently, density inversely correlates with temperature. It is conceivable that the composition of the low corona and transition region predominantly comprises such loops, variably energized by nanoflare-like heating events. In this study, we employ HYDRAD: The HYDrodynamics and RADiative emission model ~\citep{2005AGUSMSP14A..01B, 2006A&A...458..987B, 2013ApJ...770...12B, hydrad} to simulate IRIS loop under the ``hot" loop paradigm. Our analysis demonstrates that the heating profile of transition region loops cannot be adequately described by a simplistic, singular heating mechanism within such assumptions—whether the heating is spatially uniform or compact, isolated to either the footpoint or the apex of the loops. Instead, our results suggest that the heating is temporally impulsive and a spatially complex heating profile incorporating multiple heating scales is required. In addition, a forward modeling code ~\citep{2011ApJS..194...26B} is employed that can produce synthetic IRIS emission spectra along the field lines based on the output from HYDRAD. Through an extensive parameter search in the forward model, we confirm that the emitting ions are out of equilibrium and that density-dependent dielectronic recombination rates must be included to reproduce the observed line ratios. 

\section{Transition region loops observed in AR 12396 on 6th August 2015}

We present pixel-by-pixel analysis of loop brightenings observed by IRIS on 6th August 2015. The location of interest is designated as AR 12396 and the IRIS 1400\AA{} channel is the primary means for analysis since its temperature sensitivity is close to the region starting from the upper chromosphere to the lower corona containing Si IV (1393.755\AA{}, 1402.77\AA{}), O IV (1399.78\AA{}, 1401.157\AA{}, 1404.806\AA{}) and S IV (1404.808\AA{}) lines. We observed rapidly evolving several Mm long loops with transient brightenings, all exhibiting existence of sub-structure within the brightening regions which we interpret as the loops being multi-stranded (more details is available in ~\cite{2021NatAs...5..237B}). From this data, we identify six evolving loops with compact brightened masses, as shown in Figure 1, spatially centered at ($-356.0955\arcsec$, $-361.5626\arcsec$), ($-354.0993\arcsec$, $-360.5645\arcsec$), ($-352.1031\arcsec$, $-365.8877\arcsec$), ($-354.0993\arcsec$, $-376.2014\arcsec$), ($-348.1107\arcsec$, $-367.2185\arcsec$), and ($-354.0993\arcsec$, $-375.8687\arcsec$), observed at the UTC timestamps $15$:$29$:$14.610$, $15$:$31$:$40.450$, $15$:$46$:$15.520$, $16$:$11$:$46.890$, $16$:$19$:$04.420$, and $16$:$38$:$31.170$ respectively.

\begin{figure*}[t!]
\centerline{\includegraphics[width=\textwidth]{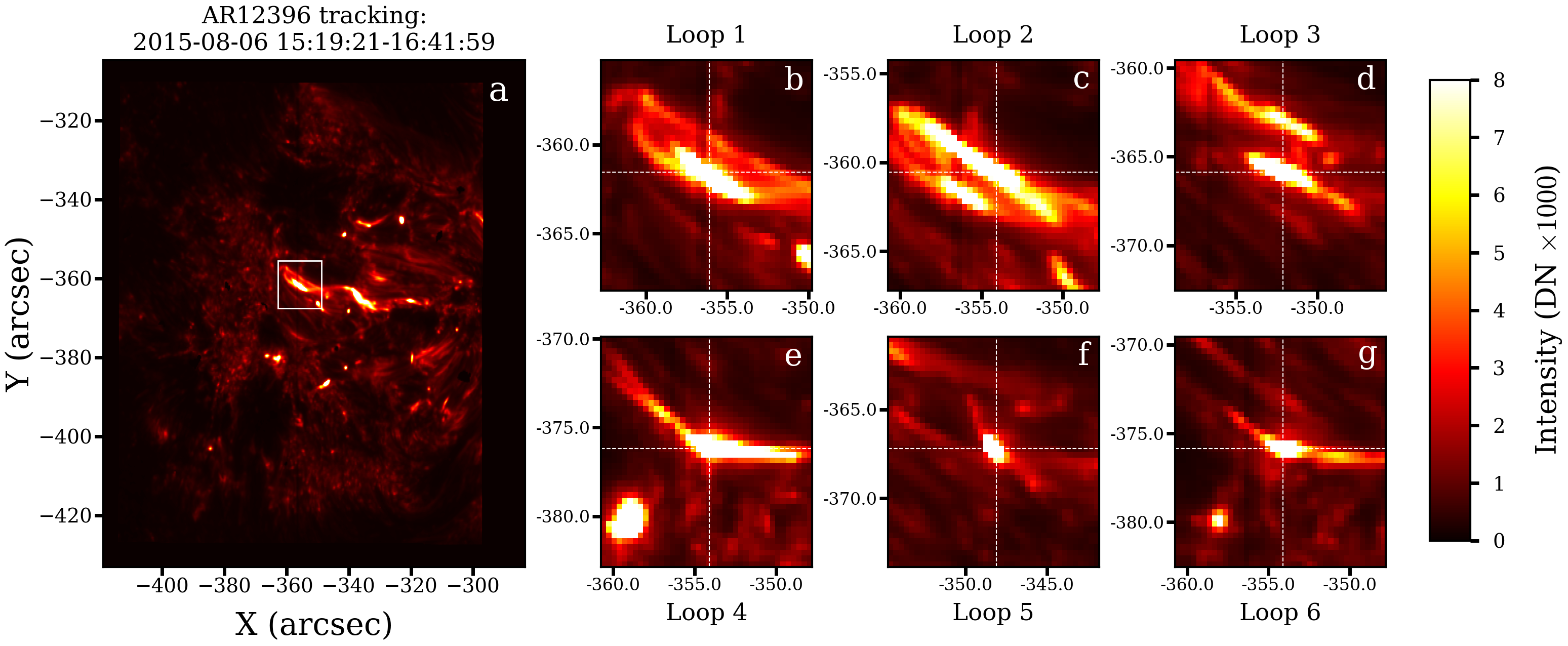}}
\caption{(a) Observations of transient loop brightenings in AR 12396 captured by the IRIS 1400Å channel on August 6, 2015. The rectangular box shows one of the candidate loop brightenings, featured more extensively in (b). Panels (b) through (g) showcase six candidate loops where transient brightenings are evident. The horizontal and vertical dashed white lines in each panel indicate the position of the most intensely brightened pixel during the onset of the brightenings.} 
\label{fig:figure1}
\end{figure*}

Spectral analysis of the Si IV 1402.77\AA{} line reveals strong Si IV 1402.77\AA{} intensity variation on the 10 - 50 second timescales during the evolution of brightening. The temporal evolution of loop dynamics extends over several minutes ($\sim 300 s$), with the initially compacted brightened mass observed at the loop apex subsequently undergoing elongation along the loop, ultimately descending on either side of the structure. During the onset of brightening, the Si IV 1402.77\AA{} line exhibits line broadening with multiple peaks having an FWHM $\Delta \lambda \approx 200$ km/s at the location of brightened region. A four-component Gaussian profile is used to fit the entire spectrum using Python routine \texttt{scipy.optimize.curve\_fit} \citep{2020SciPy-NMeth} where two Gaussian profiles are fitted for 1399.78\AA{} and 1401.157\AA{} lines and two other Gaussian profiles are fitted for the broadened 1404.806\AA{} line. 

The resultant line-of-sight velocity ($v_{LOS}$) and the non-thermal components ($v_{Non-th}$) are displayed in Figure 2. It is found that the centroids of the Gaussian profiles are located on opposite sides of the line from its rest wavelength. The Gaussian fit indicates existence of strong, asymmetric bi-directional flows with a maximum speed of more than 100 km/s toward and away from the observer as well as existence of non-thermal components as large as 150 km/s (assuming the formation temperature of O IV and Si IV lines are $10^{5.15}$ K and $10^{4.8}$ K respectively). The asymmetry of the flow may arise due to either the structure of the gravitationally stratified atmosphere or the viewing angle between the loop and the plasma flow.  However, the O IV 1401.157\AA{} line profile was only ever weakly red-shifted (maximum 25 km/s) with a single component and no significant non-thermal broadening observed.

\begin{figure}[t!]
\centerline{\includegraphics[width=0.75\linewidth]{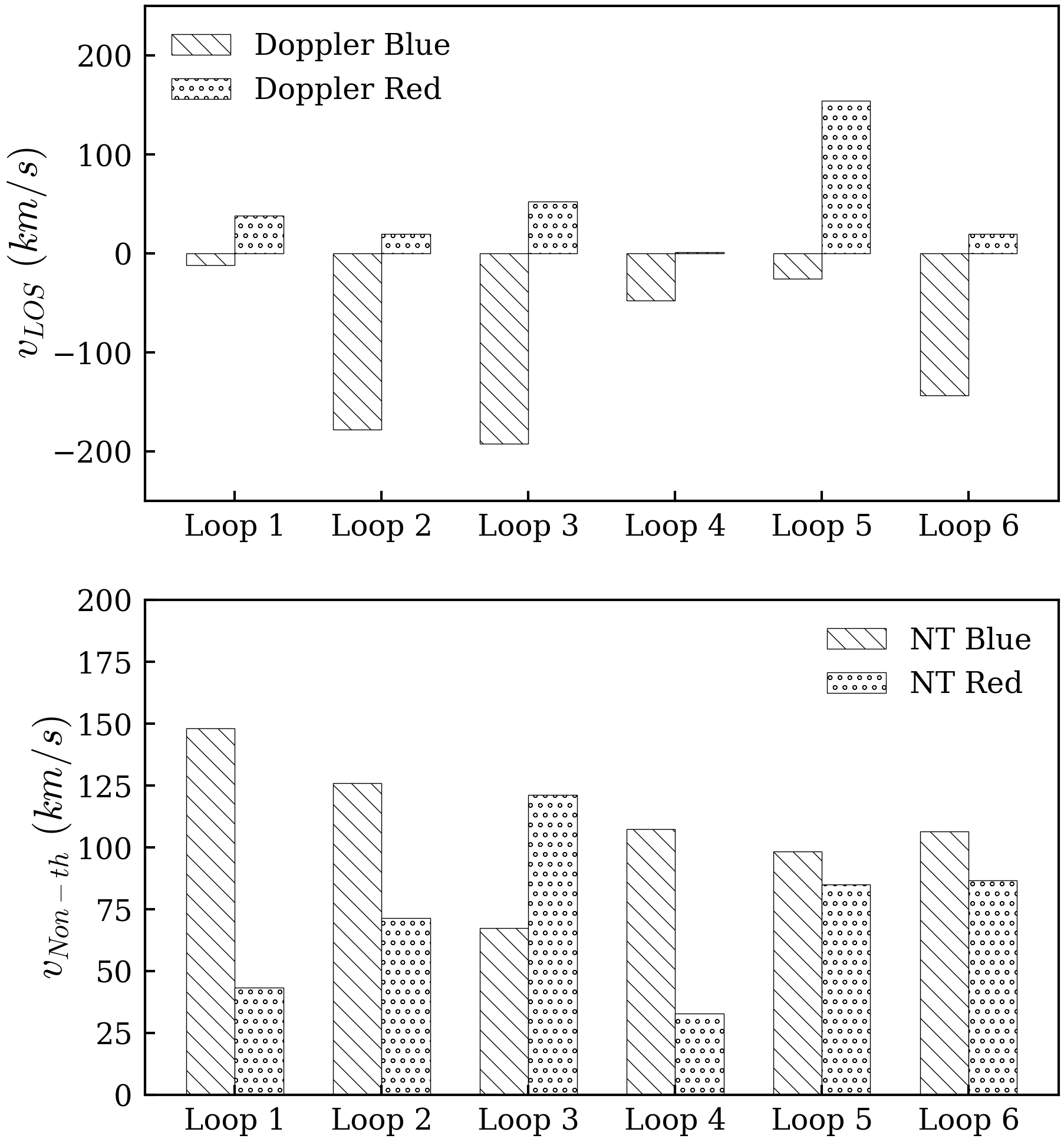}}
\caption{(Top) Bi-directional flow at the pixel exhibiting maximum brightening is determined by analyzing Doppler shifts in six identified candidate loops during the initiation of loop brightenings. The Doppler shift is computed from the mean wavelength position of the two Gaussian profiles used to fit Si IV 1402.77Å. (Bottom) The plot displays non-thermal velocity, calculated as the standard deviation of these two Gaussian profiles. The standard deviation is subsequently converted to Full Width Half Maximum (FWHM) and presented in units of $km/s$.}  
\label{fig:figure2}
\end{figure}

\begin{figure}[t!]
\centerline{\includegraphics[width=0.75\linewidth]{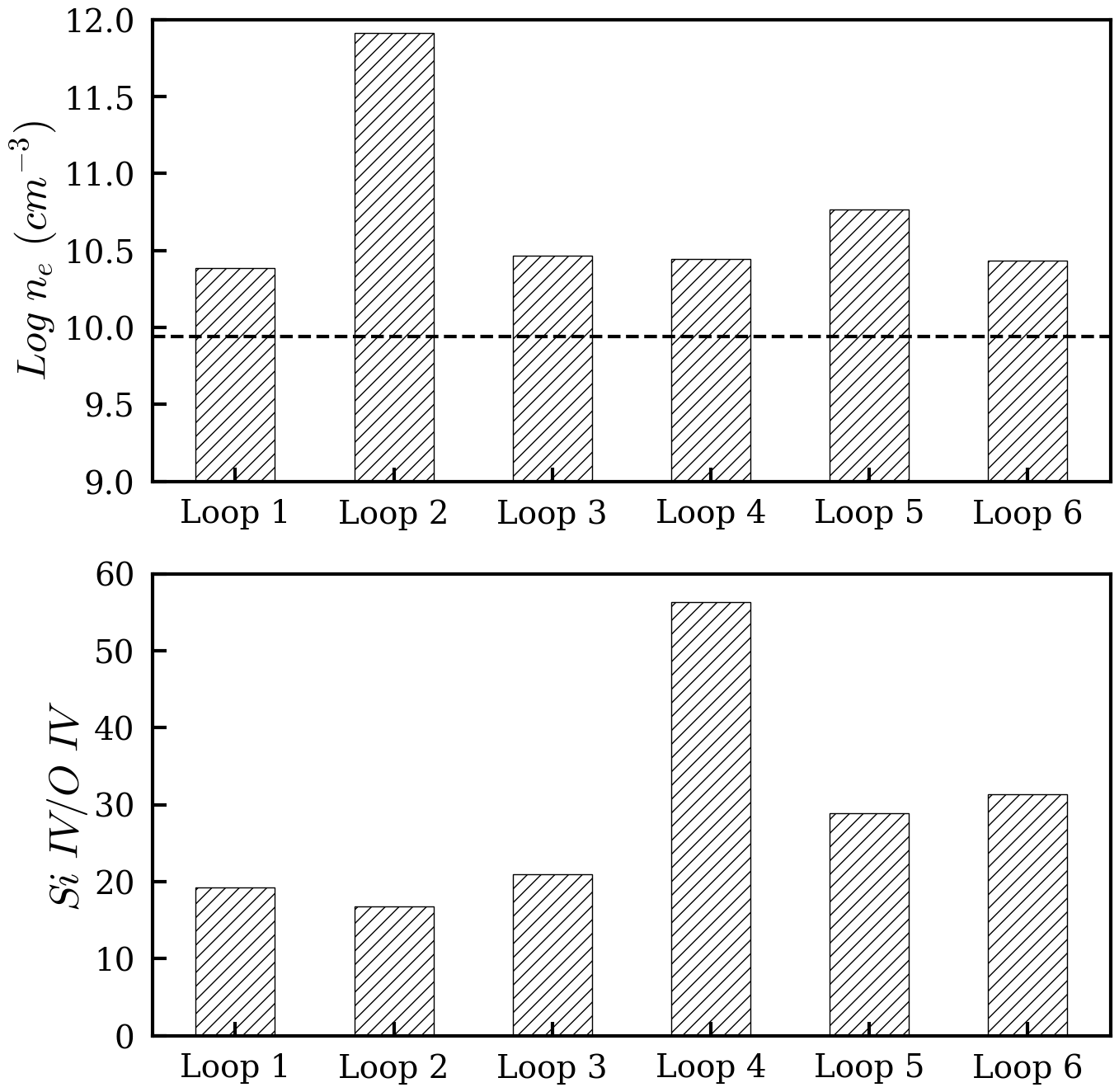}}
\caption{(Top) Logarithmic density at the location of loop brightening is computed by utilizing the ratio of O IV lines (1399.78Å and 1401.157Å) for six identified candidate loops. The dashed line indicates the average density of the non-brightening pixels ($I_{max}^{Si\, IV} < 1000$ DN). (Bottom) Concurrently, the Si IV/O IV peak ratio is determined from the fitted Gaussian profiles at those identical pixels and times where the density and velocities are computed.}  
\label{fig:figure3}
\end{figure}

The plasma density during the onset of the loop brightening is calculated from the ratio of density sensitive O IV lines (1399.78\AA{} and 1401.157\AA{}, \cite{2015arXiv150905011Y}) and is plotted in Figure \ref{fig:figure3}. Our observation reveals an increased number density ranging from $10^{10.4}$ $cm^{-3}$ to $10^{11.9}$ $cm^{-3}$ at the core of the brightening, while the average number density of these loops during the non-brightening phase remains approximately $10^{10}$ $cm^{-3}$.  Following the brightening event, the loop density reverts to the non-brightening phase values.

In addition, the ratio of the peak Si IV 1402.77\AA{} and O IV 1401.157\AA{} intensities is plotted in Figure \ref{fig:figure3}. Prior numerical simulations of transition region plasma have demonstrated that, under nanoflare-like heating conditions, the combined impact of non-equilibrium ionization (due to the impulsive nature of the heating) and density-dependent quenching of dielectronic recombination results in a pronounced enhancement in the Si IV line relative to the O IV line and thus can be used as a strong diagnostic for nanoflare heating in transition region \citep{2019ApJ...872..123B}. As shown in Figure \ref{fig:figure3}, large Si IV/O IV peak ratios (18 – 60) are found at the locations of the brightenings in the observed loops, confirming the presence of non-equilibrium ionization during the formation of emission lines.

Based on the observational findings, we come to conclusion that in order to model the observed apex brightenings with fidelity, one must satisfy several observational criteria: 1) the density profile needs to show an accumulation of dense plasma concentration ($10^{10}$ to $10^{11}$ $cm^{-3}$) at the top of the loop before the onset of the brightening, possibly via some heating mechanism. During the cooling phase, the accumulation falls down along the loop and drains into either footpoints; 2) To explain the observed brightening at the location of plasma accumulation, an impulsive heating phase (stronger than the previous one) is required. This heating mechanism is responsible for initiating the brightening and sustaining it for several hundred seconds, and would result in a bulk flow of at least 100 km/s, accompanied by non-thermal velocity components of more than 40 km/s; and 3) the Si IV/O IV peak ratio in the Si IV line has to be of order 10 or more.

\section{Modelling of Transient Heating}

\begin{table*}[]
\centering
\begin{tabular}{ccc}
\hline
            & Footpoint & Apex \\ \hline
$E_{H_0}$ ($erg/cm^3/s$)     & 0.02 -- 0.5 in interval of 0.002     & 0.05 -- 0.5 in interval of 0.01  \\
$s_0$ ($Mm$)     & 2.4, 2.8, 3.2, 3.6, 3.8, 4.2, 4.6, 5.0, 5.4, 5.8, 6.2, 6.6     & 6.8, 7.2, 7.6  \\
$s_h$ ($Mm$)     & 0.25, 0.50, 0.75, 1.0, 1.25, 1.50, 1.75, 2.0, 2.25, 2.5        & 0.1, 0.2, 0.3, 0.4, 0.5  \\
$t_0$ ($s$)      & 0                                                              & 150, 160, 170, 180, 190, 200, 210, 220, 230, 240, 250  \\
$\Delta t$ ($s$) & 2, 4, 6, 8, 10                                                 & 1, 2, 3, 4, 5, 6    \\
$t_r$ ($s$)      & 1, 2, 3, 4, 5                                                  & 0.5, 1, 1.5, 2, 2.5, 3    \\
$t_d$ ($s$)      & 1, 2, 3, 4, 5                                                  & 0.5, 1, 1.5, 2, 2.5, 3    \\
$\delta$ ($s$)   & 5, 10, 15, 20, 25, 30                                          & 10, 20, 30, 40, 50, 60   \\
$n$        & 5, 10, 15, 20, 25, 30, 35, 40                                  & 1, 2, 3, 4, 5, 6, 7, 8    \\ \hline
\end{tabular}
\caption{Parameter space for HYDRAD simulations of transition region hot loops observed by IRIS. $E_{H_0}$ represents the maximum volumetric heating rate, $s_h$ and $s_0$ denote the spatial scale and the center of the heating profile, respectively. $t_0$ indicates the onset of the heating (relative to the start of the simulation clock), $\Delta t$ indicates the heating duration, $t_r$ and $t_d$ refer to the rise and decay time of the temporal heating profile, $\delta$ denotes the period of the temporal profile for each heating event, and $n$ represents the number of heating events.}
\label{tab:Table 1}
\end{table*}

To effectively reproduce the spectra from intermittent brightenings with significant line broadenings, our loop model integrates a radiation treatment capable of accommodating radiative emission from non-equilibrium ion populations. Additionally, since the location of the heating characteristically influences the hydrodynamic evolution of the loop and its radiative balances; we treat this parameter a free variable in the model and explore the parameter space through numerical simulations. Both numerical and observational evidence suggests that coronal loops, including those with smaller scale heights, can be impulsively heated either at their footpoints or uniformly throughout the loop, regardless of the specific heating mechanism. 

An exhaustive search in the parameter space for both footpoint heating and uniform heating of the loop is conducted. However, none of the models were able to reproduce all of the spectral features that are ubiquitous in the observed loops, indicating the need for a bottom-up model-building approach. We address this problem by introducing increasingly complex heating profile guided closely by observational data, moving beyond previous assumptions. 

First, the condensation is achieved by gently heating the footpoint of the loops with a train of intermittent chromospheric heating. Once a condensation of the observed plasma density is reached ($10^{10}$ to $10^{11}$ $cm^{-3}$) at the top of the loop, a new train of stronger, impulsive heating is introduced at the site of condensation. This initiates the strong nanoflare-like heating (magnetic reconnection) and produces a bi-directional velocity component near the heating site, emulating thermal line broadening found in observation. In order to reproduce the large Si IV/O IV peak ratio, we investigated both equilibrium and non-equilibrium ionization populations and examined the effects of keeping the dielectronic recombination rates constant for low-density coronal plasma, as well as allowing for variation in the rates for high-density transition region plasma. Below we outline the physical characteristics and spatiotemporal heating profile of the model loop. 

The simulated loop has a length of 10 Mm along with each footpoint having 2.2 Mm buried in the chromosphere resulting the total length to be $L =$ 14.4 Mm. The spatial profile of the heating is a Gaussian distribution with spatial width $s_H$ along the loop, centered at the location $s_0$, and spreaded across the field line. Thus the volumetric heating rate can be expressed as, $E_H(s) = E_{H_0} e ^ {-(s-s_0)^2/2s_{H}^2}$, where $E_{H_0}$ is the maximum volumetric heating rate. The equation is flexible enough to localize the heating and place it anywhere along the loop including the case of uniform heating where $s_{H} \to \infty$. In the present case, we introduce a temporally triangular-shaped heating profile (characterized by $\Delta t$ heating duration and $\delta$ period) with linear rise ($t_r$) and decay ($t_d$). Spatially, we adopt a broad Gaussian profile to heat the footpoints of the loop, with relatively mild volumetric heating near the footpoints. This initiates formation of a condensation at the apex of the loop. Once a condensation is formed, the apex is heated by a stronger, sharp volumetric heating having a narrow Gaussian profile concentrated at the location of plasma accumulation. In order to closely align simulation outcomes with observational data, several asymmetries are integrated into the modeling approach: (1) a $10\%$ asymmetry in heating magnitude between the two footpoints, and (2) a spatial deviation of $3\%$ in the location of apex heating from the center of the loop ($s_0=L/2+0.03L$). Consequently, the second criterion dictates the direction of draining of the mass concentration through the footpoints. However, observations indicate mass drainage at both footpoints, albeit with one footpoint exhibiting greater drainage than the other. To replicate this phenomenon, multiple loops are simulated, each featuring a random spatial placement of apex heating within a $3\%$ variation from the center of the apex ($s_0=L/2 \pm 0.03L$). Synthesizing spectra from the image timeseries of these simulated loops recreates loop dynamics characterized by predominant draining through one footpoint, with a comparatively weaker flow observed at the other footpoint. In Table \ref{tab:Table 1}, we present the parameter space associated with each variable related to the heating profile.

Following an extensive search through the parameter space, we have determined that the values of the variables listed in Table \ref{tab:Table 2} best reproduce the observed loop. Under the optimal conditions, the loop remains in a steady-state with a temperature below $10^5$ K and an apex density of $10^{9}$ $cm^{-3}$ satisfying a hot-loop solution. With a gentle heating in the footpoint, characterized by a heating rate of 0.018 $erg/cm^3/s$ with $s_0=5.8$ Mm and $s_H=1$ Mm, a condensation of $n_e = 10^{10.3}$ $cm^{-3}$ is seen to be gradually formed over a span of nearly 100 seconds. Similar condensations are observed during thermal-non-equilibrium cycles in coronal loops ~\citep[e.g.,][]{2017ApJ...835..272F, 2018ApJ...855...52F, 2020PPCF...62a4016A}. Upon reaching a steady condensation state at t = 190 s, a short, isolated nanoflare with a heating rate of 0.3 $erg/cm^3/s$ lasting for 4 seconds is introduced, mimicking observed impulsive energy release due to reconnection. This sudden heating elevates the temperature of the surrounding plasma to coronal temperature, and a strong bi-directional flow ($> 50$ km/s) along the loop arise. Forward modelling reveals that the heated plasma is radiating with 400 DN/s when the effects of non-equilibrium ionization and density-dependent quenching of dielectronic recombination are accounted for. Incorporating the observed non-thermal line broadening ($\sim 50$ km/s) into the synthetic Doppler shift reproduces the observed radial bi-directional flow of $100$ km/s. Through parameter space exploration, our investigation reveals that sustaining simulated brightenings equivalent to observed duration ($> 100 s$) necessitates a series of nanoflares ($n = 4$), as opposed to a single nanoflare event, regardless of its temporal span. This requirement holds true while ensuring that the heating is sufficiently rapid to maintain non-equilibrium ionization. After the heating is turned off, the hot mass content drains to the chromosphere during the cooling phase of the loop. 

\begin{table}[b]
\centering
\begin{tabular}{cccc}
\hline
    & \begin{tabular}[c]{@{}c@{}}Left\\ Footpoint\end{tabular} & \begin{tabular}[c]{@{}c@{}}Right\\ Footpoint\end{tabular} & Apex \\ \hline
$E_{H_0}$ ($erg/cm^3/s$)  & 0.018  & 0.018  & 0.3  \\
$s_0$ ($Mm$)  & 5.8  & 13  & 7.2  \\
$s_h$ ($Mm$)  & 1.0  & 1.0  & 0.2  \\
$t_0$ ($s$)  & 0  & 0  & 190  \\
$\Delta t$ ($s$)  & 4  & 4  & 4    \\
$t_r$ ($s$)  & 2  & 2  & 2    \\
$t_d$ ($s$)  & 2  & 2  & 2    \\
$\delta$ ($s$)  & 10  & 10  & 40   \\ 
$n$  & 20  & 20  & 4    \\ \hline

\end{tabular}
\caption{Optimal parameters for simulating the transition region hot loops to closely replicate the IRIS observation.}
\label{tab:Table 2}
\end{table}

\begin{table*}[t!]
\centering
\begin{tabular}{ccccc}
\hline
                                                                            & \begin{tabular}[c]{@{}c@{}}Density\\ log $n_e$ ($cm^{-3}$)\end{tabular} & \begin{tabular}[c]{@{}c@{}}Velocity\\ (km/s)\end{tabular} & \begin{tabular}[c]{@{}c@{}}Intensity\\ (DN/s)\end{tabular} & Si IV/O IV \\ \hline
Observations from IRIS                                                      & 10.4-11.9                                                       & 50-200                                                    & 180-700                                                    & 18 – 60           \\
\begin{tabular}[c]{@{}c@{}}Equilibrium (Low Density Limit)\end{tabular}           & 10.5-11                                                         & \textgreater 50                                          & 68                                                         & 2               \\
\begin{tabular}[c]{@{}c@{}}Non-Equilibrium (Low Density Limit)\end{tabular}       & 10.5-11                                                         & \textgreater 50                                          & 80                                                         & 3               \\
\begin{tabular}[c]{@{}c@{}}Equilibrium (Density Dependent Rates)\end{tabular}     & 9.8-10.5                                                        & \textgreater 50                                          & 200                                                        & 10              \\
\begin{tabular}[c]{@{}c@{}}Non-Equilibrium (Density Dependent Rates)\end{tabular} & 10-10.7                                                         & \textgreater 50                                          & 400                                                        & 30              \\ \hline
\end{tabular}
\caption{Comparison of observable diagnostics between the spectra of observed loop and the spectra of simulated hot loop in solar transition region.}
\label{tab:Table 3}
\end{table*}

\begin{figure}[t!]
\centerline{\includegraphics[width=0.75\linewidth]{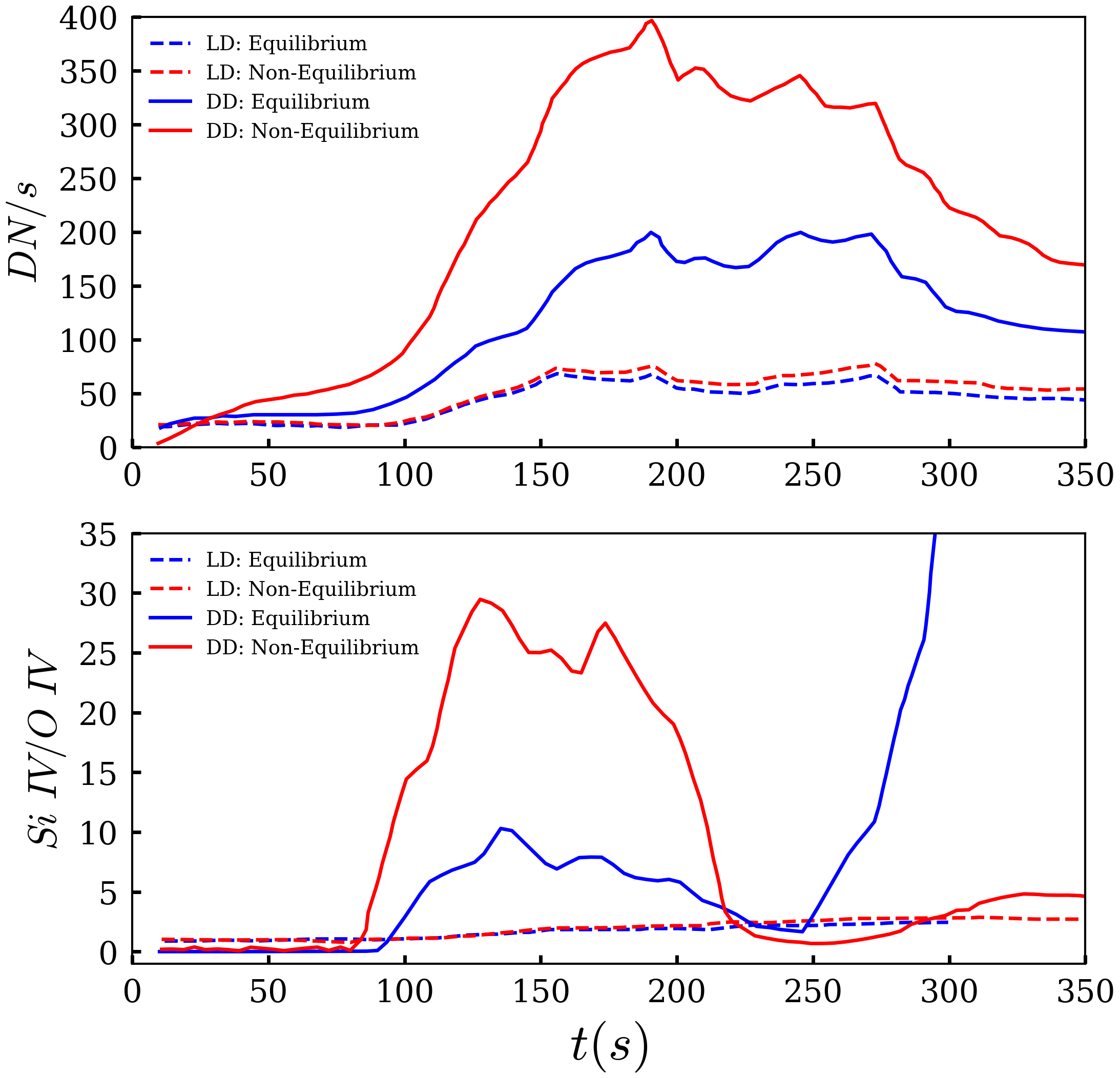}}
\caption{(Top) Si IV intensity and (Bottom) Si IV/O IV peak ratio derived from the synthetic spectrum simulated by the Forward Modeling code. The results from non-equilibrium (NE) ionization including the density dependent dielectronic recombination rates (LD = low density limit and DD = density-dependent rates) produce stronger pixel intensity and larger Si IV/O IV peak ratio, aligning quantitatively with observations.}  
\label{fig:figure4}
\end{figure}

In our subsequent analysis, we explore various criteria for ion populations to investigate the intensity and evolution of synthesized Si IV and the Si IV/O IV peak ratio at the apex. As depicted in Figure \ref{fig:figure4}, it is evident that employing non-equilibrium ionization criteria yields a stronger Si IV emission profile, accompanied by a notably larger Si IV/O IV peak ratio. To elucidate this result, we underscore the optimal heating parameters in Table \ref{tab:Table 2}, characterized by the apex heating of 0.3 $erg/cm^3/s$ for a duration of 4 seconds. This heating timescale is fast enough to induce a departure from equilibrium ionization state, leading to an enhanced population of Si IV. Concurrently, the rapid increase of plasma pressure due to the nanoflare heating drives the transition region plasma deeper into the atmosphere. In this high-density regime, the density-dependent quenching of dielectronic recombination becomes significant, thereby extending the lifetime of the enhanced Si IV population. Comparative analysis, presented in Table \ref{tab:Table 3} and Figure \ref{fig:figure4}, confirms this explanation as the incorporation of density-dependent dielectronic recombination rates results stronger quantitative agreement against the observed data in terms of intensity, density, and Si IV/O IV peak ratio.

Thus, the high Si IV/O IV ratio in the IRIS data can be used as a strong diagnostic tool to characterize the ionization state of the population and to identify the events of impulsive heating. This proves the fidelity of forward model that includes non-equilibrium ionization conditions and accommodates recombination rates over a large variation of density during the calculation of ion population and the emission line intensities. It is essential to exercise caution when assessing number density derived from line ratios, as many density diagnostic methods assume that the emitting plasma is in both thermal and ionization equilibrium (e.g., \cite{2015arXiv150905011Y}). As for an example, if the observed density is computed from the density diagnostics prescribed by CHIANTI's O IV line ratios \citep{2015A&A...582A..56D}, it is imperative that the modeled density be computed using identical line ratios (obtained from forward-modeled spectra) and derived from the same atomic database.

\subsection{Effects of Magnetic Field}

In the observation, large non-thermal broadenings are evident in both Si IV and S IV lines, whereas lighter ions like O IV display no such attributes. Such preferential ion line broadening mechanism presents a challenge for HYDRAD simulations and prompts investigation into the multi-species behavior of plasma under the influence of magnetic fields. One plausible explanation involves ion cyclotron instabilities during magnetic reconnection, wherein ion heating rates exhibit a weak proportionality to the ion mass~\citep[e.g.,][]{1985JGR....9012209S, 1993JATP...55..647F, 2003NPGeo..10..101M}, and critical drift velocities are lower for the more abundant heavier ions. For a detail discussion on how ion cyclotron instability energize different ion species and their charge states, we recommend our previous literature~\cite{2021NatAs...5..237B}.

An intriguing finding from this study is the requirement of density enhancement at the location of brightening.  In the simulation, this is accomplished by inducing plasma condensation at the loop apex through gentle, periodic heating at the footpoint, resembling the so-called thermal non-equilibrium heating ~\citep[e.g.,][]{2020PPCF...62a4016A}. As detailed in our previous study~\citep{2021NatAs...5..237B}, the occurrence of brightening phenomena coincides when regions of density enhancement experiences field line braiding, possibly triggering trains of magnetic reconnection events. If the ``hot" loop framework accurately represents these loops, it implies that density enhancement is closely associated with the onset of impulsive bursts of reconnection. Given the high conductivity of the transition region plasma (characterized by $S = 10^7$ assuming $B = 100$ $G$, $n = 10^{11}$ $cm^{-3}$, and $\eta = 10^8$ $m^2/s$), numerous plasmoids with secondary instabilities are expected to exist in these loops \citep{2015ApJ...813...86I}. Under such conditions, a density enhancement would cause the secondary current sheets to become thinner, potentially leading to higher current density at the X-point \citep[see]{2012APS..DPPBI2005H, 2013PhPl...20f1206N}. While a precise description of such reconnection mechanism remains elusive without an MHD model, the observation of a nearly order-of-magnitude increase in density, coupled with highly impulsive heating behavior, suggests the likely involvement of smaller-scale, sub-resolution structures in the heating process.

Alternatively, transition region loops may conform more closely to the ``cool" loop framework \citep[e.g.,][and references therein]{1979A&A....77..233H, 1982A&A...112..366M, 1986ApJ...301..440A, 1988ApJ...328..334K, klimchuk1992static}. In this scenario, temperature gradients within the loops are generally shallow, with slight pressure stratification and relatively denser plasma ($n_e \gtrsim 10^{10}$ $cm^{-3}$). Thermal conduction plays a minor role in such loops, as there is local equilibrium between radiation and energy throughout. Due to their structure and energy balance, ``cool" loops exhibit stronger emissions in transition region lines ($\sim 10^5$ K). It is possible that under the ``cool" loop framework, the need for gentle, periodic footpoint heating to produce density enhancement in the loop apex may be alleviated, as the steady-state density of the loops could be sufficiently high to drive heating under the "high-density" limit. Although this study does not delve into modeling the observed IRIS loops within the ``cool" loop paradigm, we intend to explore this aspect in our forthcoming literature.

\section{Conclusion}

HYDRAD modeling of transient brightenings in the solar transition region loops, observed in IRIS 1400 Å slit-jaw images, reveals compelling evidence that these loops are subject to impulsive heating and is likely the result of a complex heating profile that incorporates multiple heating scales. Systematic exploration across a wide parameter space of ``hot" loop paradigm confirms the inability to replicate the observed dynamics of intensity variation, irrespective of the degree of fine-tuning, using either simple footpoint heating or apex heating. The study underscores the necessity of introducing heating at various locations and corresponding spatial scales to accurately reproduce the observed intensity profiles.

Impulsive heating events such as these are widespread in both the chromosphere and the transition region and the model indicates that the heating mechanisms drive the ion population out of equilibrium, leading to large intensity ratios between selective ions and strong non-thermal line broadening. The spectral analysis we present, that accounts non-equilibrium ionization and incorporates density dependent dielectronic recombination rates, provides a quantifiable observational diagnostics to critically examine these heating events and their spectral signatures. However, the precise origin of the preferential ion heating observed in these brightenings cannot be answered by field-aligned hydrodynamic and radiative simulations alone, and requires MHD+kinetic analysis on a case-by-case basis. This has been carefully investigated in observational literature and possible underlying mechanisms, including nanoflares, turbulence, plasmoid instabilities, and magneto-acoustic shocks are proposed~\citep[e.g.,][and references therein]{2014Sci...346A.315T, 2014Sci...346B.315T, 2014Sci...346D.315D, 2015ApJ...813...86I, 2015ApJ...799L..12D, 2018ApJ...868L..27P, 2021NatAs...5..237B, 2021cosp...43E1798A, 2021SoPh..296...84D}. Nonetheless, it is plausible that small-scale, fast impulsive heating events operate throughout the solar upper atmosphere; creating the quiet Sun, quiescent active regions, and perhaps even playing a role in energizing and accelerating the solar wind.

\section{Acknowledgments}
The authors extend their gratitude to James A. Klimchuk for his insightful input regarding the density enhancement within the hot and cool loop framework. This research was partially supported by the National Aeronautics and Space Administration under Award Number NNX15AF97A and the National Science Foundation under Award Number 2206589. We acknowledge the contributions of the LMSAL/IRIS science team and their support for the IRIS data search portal.

%

\vspace{5mm}
\facilities{Interface Region Imaging Spectrograph (Slit-Jaw Imager Si 1400\AA{}, Spectrograph FUV passband 1389\AA{} to 1407\AA{})}


\software{astropy \citep{2013A&A...558A..33A, 2018AJ....156..123A},  
          SciPy \citep{2020SciPy-NMeth}, 
          HYDRAD \citep{hydrad}
          Forward Modelling Code \citep{2011ApJS..194...26B}
          }



\bibliography{sample631}{}
\bibliographystyle{aasjournal}



\end{document}